\documentclass{llncs}

\usepackage[utf8]{inputenc}
\usepackage{amsmath}
\usepackage{amsfonts}
\usepackage{amssymb}
\usepackage{graphicx}
\usepackage{url}
\usepackage{authblk}
\usepackage{color}
\usepackage{booktabs}

\usepackage{times}     
\usepackage[final]{microtype}  

\def \calC {\mathcal{C}}
\def \calU {\mathcal{U}}

\DeclareMathOperator*{\argmax}{arg\,max}

\title{The City Pulse of Buenos Aires}

\author{Carlos Sarraute\inst{1} \and
Carolina Lang\inst{1} \and
Nicolas B. Ponieman\inst{1} \and
Sebastian Anapolsky\inst{2}
}

\institute{Grandata Labs, Argentina
\and Mobility and transport specialist
}

\begin{document}
\pagestyle{plain}

\maketitle

\section{Introduction}

Cell phone technology generates massive amounts of data. Although this data has been gathered for billing and logging purposes, today it has a much higher value, because its volume makes it very useful for big data analyses. 
In this project, we analyze the viability of using cell phone records to lower the cost of urban and transportation planning, in particular, to find out how people travel in a specific city (in this case, Buenos Aires, in Argentina).
We use anonymized cell phone data to estimate the distribution of the population in the city using different periods of time. 
We compare those results with traditional methods (urban polling) using data from Buenos Aires origin-destination surveys.
Traditional polling methods have a much smaller sample, in the order of tens of thousands (or even less for smaller cities), to maintain reasonable costs.
Furthermore, these studies are performed at most once per decade, in the best cases, in Argentina and many other countries.
Our objective is to prove that new methods based on cell phone data are reliable,
and can be used indirectly to keep a real-time track of the flow of people among different parts of a city.
We also go further to explore new possibilities opened by these methods.


\section{Mobile Data Source}

We applied our methodology to Buenos Aires city, the capital of Argentina, which has 2,890,151 inhabitants \cite{censo2010} and is the main political, financial and cultural center of the country.
Buenos Aires city is formally divided in 48 neighborhoods, which are grouped for political and administrative purposes in 15 communes.

We used a dataset of geolocalized CDR (call detail records), from which we examine the mobility patterns of mobile phone users.
The dataset is anonymized, and contains no personal information such as name or actual phone number. The users' numbers were hashed with a cryptographic function during the generation of the dataset. 

The high penetration of cell phone technology
in the city allows us to estimate the mobility patterns of all the inhabitants from this data.
Using mobile phone records to understand users mobility is an active area of 
research~\cite{gonzalez2008understanding,song2010limits,ponieman2013,Ponieman2015mobility},
and has applications in diverse areas such as urban planning~\cite{wang2012understanding}, epidemiology~\cite{wesolowski2015impact,sarraute2015descubriendo} and disaster recovery~\cite{lu2012predictability}.

In this work, the studied dataset contains about 4.95 million mobile phone users (1,000 times the number of people in the Buenos Aires survey~\cite{enmodo2010});
it also contains more than 200 million call records generated by these users during a period of five months
(from November 1st, 2011 to March 30th, 2012).
Each record contains the origin (hashed number of the caller), destination (hashed number of the callee), timestamp, duration of the call and antenna used to connect. In addition, we have the geolocalization of the antennas.
We used that information to map the antennas to a certain commune, and we used the map [call$\rightarrow$antenna] as dataset of geolocated calls.

\section{Methodology}

In this section we explain the methodology we used to adapt the Call Detail Records (CDRs) to our objective.

The first step of our method generates, for each particular user, a \textit{Location Distribution Matrix} (LDM) that shows the probability of the user being in a commune $c$ at a given time $t$ of the week.
The second step defines a criteria to consider only users whose LDMs give us enough information about their mobility patterns.
The last step scales our sample using the population values from the census data.

\subsection{Generation of Location Distribution Matrices}
\label{sec:LDMs}

We separated a typical week into four day groups and four hour groups, as shown in  Table~\ref{table:weekSlots}.

\begin{table}[th]
	\vspace{-0.2cm}
	\caption{Day groups and hour groups used in our analysis}
	\label{table:weekSlots}
\begin{minipage}{0.36 \linewidth}	
	\centering
	\begin{tabular}{ l } 
		\toprule 
		\textbf{Day groups}	\\
		\midrule 
		Monday to Thursday	\\
		Friday				\\
		Saturday			    \\
		Sunday				\\
		\bottomrule 
	\end{tabular}
\end{minipage}
\begin{minipage}{0.63 \linewidth}	
	\centering
	\begin{tabular}{ l l } 
		\toprule 
		\textbf{Hour groups}			\\
		\midrule 
		Morning & 5am - 11am					\\
		Noon    & 11am - 3pm						\\
		Afternoon & 3pm - 8pm					\\
		Night   & 8pm - 5am (of next day)	\\
		\bottomrule 
	\end{tabular}
\end{minipage}

\end{table}

This selection is based on the fact that Monday to Thursday are typical working days, Fridays show different mobility patterns (specially at night), and weekends present a completely different pattern.

The hour group selection corresponds to an analysis realized with the data of 
\cite{enmodo2010}, from which we determined the peaks and valleys of mobility, for a typical working day in the city.

Let $\calC$ be the set of communes and $R_{u,d,h,c}$ the number of calls made by user $u$ on day group $d$, hour group $h$, in commune $c$.
The proportion of calls (i.e., the cell values of the LDM) that a certain user $u$ made in commune $c$
during a combination of day group $d$ and hour group $h$ is 
$$P_{u,d,h,c} = \frac{R_{u,d,h,c}}{\sum_{c' \in \calC}{R_{u,d,h,c'}}}$$ 
or $0$ if the denominator is zero. 
The matrix $P_u$ is the \emph{Location Distribution Matrix} of user $u$.

\subsection{Criteria for Filtering Users}
\label{sec:filterUsers}

We filter the users that don't provide enough information on their location;
more precisely we only take into account the users that have enough calls in every one of the 16 day/hour groups.
That is, the user $u$ is kept if 
$$\sum_{c' \in \calC}{R_{u,d,h,c'}} \geq \tau$$ 
for any combination of $d$ and $h$, given a threshold $\tau$ (in our study $\tau = 1$).
After filtering, we obtain a set of 73,000 users which we denote $\calU$.

\subsection{Scaling up to Census Population}
\label{sec:scaling}

First, we determine the home commune $H_u$ for every user $u \in \calU$.
We consider that a user is at home on weekdays, at night:
$$ H_u = \argmax_{c \in \calC}  R_{u,\text{weekday},\text{night},c}  $$
In case of a tie, we decide randomly. We registered only 395 ties among the set of valid users $\calU$
(0.56\% of the cases).

With that information, we extend our predictions using the census data \cite{censo2010}. 
The scaling factor $F_c$ for commune $c$ is:
$$ F_c = \dfrac{ \text{pop}_c}{ \# \{ u \in \calU | H_u = c \}  } $$ 
where $\text{pop}_c$ is the population of commune $c$ according to the census.
The range of scaling factors goes from $17.26$ in commune 2 to $93.29$ in commune 8.

We now define the expected quantity of people in a commune $c$, 
during a combination of day group $d$ and hour group $h$,
in terms of the proportion of calls of each user in $c$ and the scaling factor of their home commune:
$$ EP_{d,h}[c] = \sum_{u \in \calU} \left( P_{u,d,h,c} \cdot F_{H_u} \right) $$
Additionally, the expected quantity of people found in commune $c$, during a day group $d$ and hour group $h$, and that live in commune $c'$ is given by:
$$ EP_{d,h}[c][c'] = \sum_{u \in \calU | H_u = c'} \left( P_{u,d,h,c} \cdot F_{c'} \right) . $$
Note that $EP_{d,h}[c] = \sum_{c' \in \calC} EP_{d,h}[c][c']$.
Having presented the methodology, we now describe the results obtained.


\section{Results}

\subsection{Validation Against the Survey}
\label{sec:validation}

We first validated the proposed methodology, by comparing it with the most traditional method used among the urban mobility studies in Buenos Aires: 
the origin-destination survey \cite{enmodo2010,anapolsky2013}.

\begin{figure}[thb]
	\centering
	\includegraphics[width = \columnwidth]{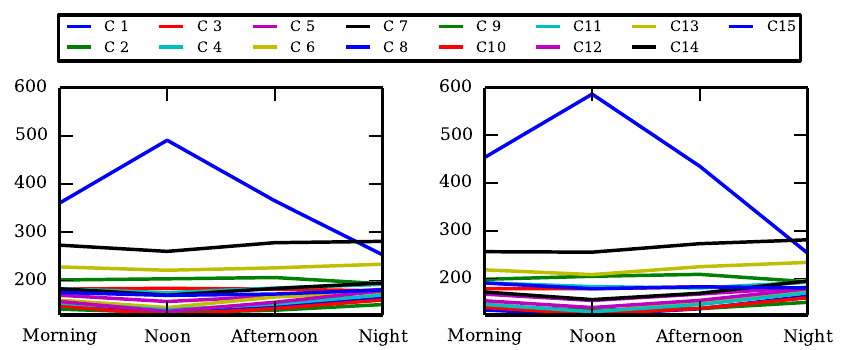}
	\caption{Comparison between the ENMODO survey (left) and our analysis (right), for a typical working day, and for all the communes in Buenos Aires. The numbers in the legend correspond to the commune numbers. The $y$ axis shows the estimations in thousands of people.}
	\label{fig:enmodoVsTelco}
	\vspace{-0.2cm}
\end{figure}

In Fig.~\ref{fig:enmodoVsTelco}, we see that the results obtained are similar (both plots show the same growth patterns for each commune). 
A more detailed analysis of the differences between the two data sources shows
that the average difference is 5\%.
The highest variation appears in Commune 1 (20\% in the morning hour group) and the second highest in Commune 6 (11\% in the noon hour group). 
For a more detailed analysis, we refer the reader to \cite{anapolsky2014clatpu}.

\subsection{Extension to Weekends}

Given that we have successfully validated our proxy methodology with the origin-destination survey, we can now use it to extend the analysis to other time periods. 
We examine here the mobility during the weekends.
The mobility survey \cite{enmodo2010} does not include this information;
we are thus presenting here new results on the mobility of the citizens of Buenos Aires. 

\begin{figure}[thb]
	\centering
	\includegraphics[width = \columnwidth]{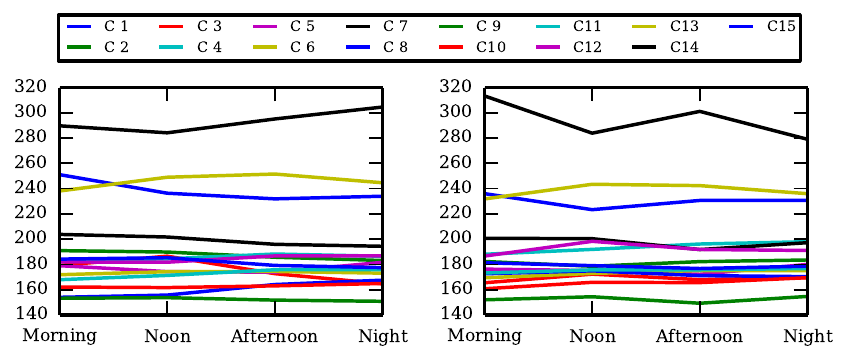}
	\caption{Predictions for a typical Saturday (left) and Sunday (right) according to our methodology, for all the communes in Buenos Aires.
		The numbers in the legend correspond to the commune numbers and the $y$ axis shows the estimations in thousands of people.}
	\label{fig:allCommunesWeekends}
	\vspace{-0.2cm}
\end{figure}

The patterns for weekends (Fig.~\ref{fig:allCommunesWeekends}) are very different: Commune 1, the central business district of the city, is not a major pole of attraction (as it is during weekdays), whereas other communes (mainly Commune 14) are more attractive for citizens on weekends. Commune 14 is well known for its bars, restaurants and night clubs, so this pattern coincides with our insight on the social life in this commune. 

\subsection{Analysis of Commune 14}
\label{sec:palermo}

\begin{figure}[t]
	\centering
	\includegraphics[width = 0.60 \linewidth, clip=true, trim=14 1 15 9]{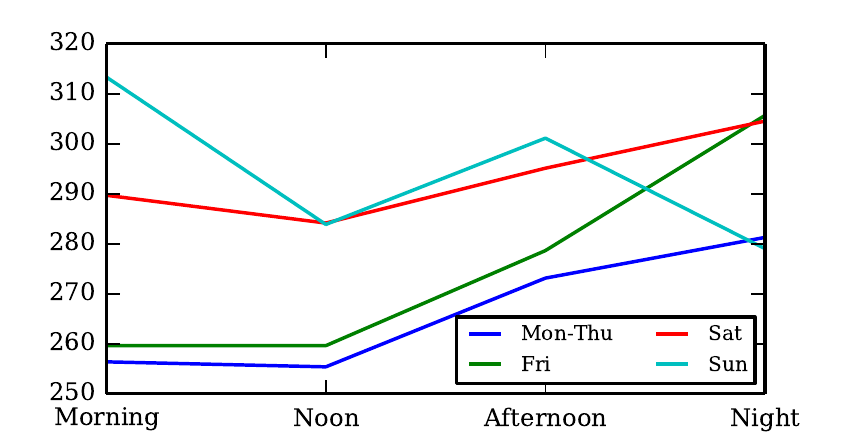}
	\caption{Predictions according to our method, for the different day types, for Commune 14 (Palermo). The $y$ axis shows the estimations in thousands of people.}
	\label{fig:palermo}
	\vspace{-0.2cm}
\end{figure}

We analyse in more detail Commune 14 (Palermo), which has very particular characteristics (see Fig.~\ref{fig:palermo}).
First of all, we remark it has a typical residential commune pattern for weekdays (with a lower concentration of people during working hours, and a higher concentration at night). During weekends, however, Commune 14 shows a special behavior due to its role as social and nightlife hub. 
During Fridays, we notice an increase of people during the night when compared to other weekdays, which we attribute to people going out. Saturdays show an increase in population across all time groups, with a peak at night that is similar to the one on Friday, and Sunday night has the same quantity of people than a regular working day at night, probably because people will have to go to work on the following day. Moreover, we notice a similar number of people on Friday night compared with Saturday morning, and on Saturday night compared with Sunday morning. This fact may be explained considering nightlife in Buenos Aires extends into the morning (even until 8am).
All these observations are coherent with our knowledge of the city.

\subsection{The City Pulse Matrix}
\label{sec:ODMatrix}

\begin{figure}[th]
	\centering
	{\includegraphics[width = 0.5\columnwidth, clip=true, trim=18 5 26 34]{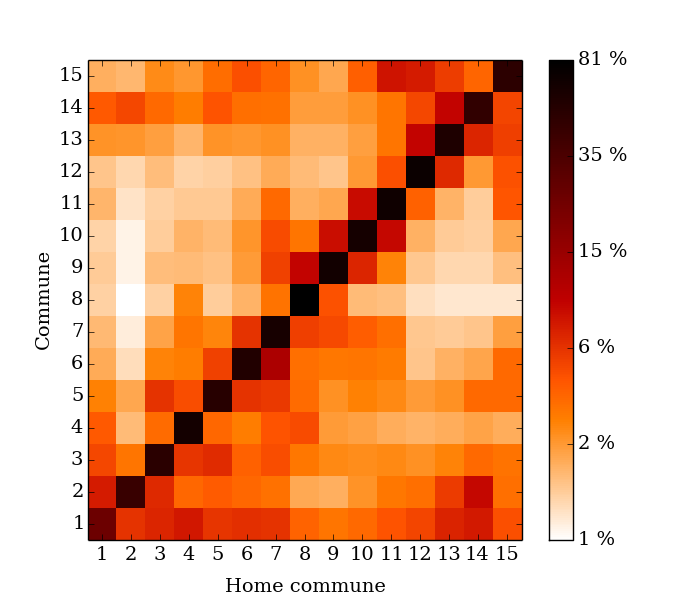}}
	\caption{Visualization of the \emph{City Pulse Matrix} generated with our methodology, for a weekday (Monday to Thursday) noon, with values normalized by row.}
	\label{fig:CPMatrix}
\end{figure}

The urban mobility information can be used to generate what we call the
\emph{City Pulse Matrix} (CPM), a 2-dimensional matrix such that, for any day group $d$ and hour group $h$, 
$${CPM}[i][j] = EP_{d,h}[i][j] .$$
Fig.~\ref{fig:CPMatrix} shows our visualization of the matrix generated by our predictions, on a typical weekday noon (which is the time period that varies the most with respect to weekday nights).

We can see in Fig.~\ref{fig:CPMatrix} that there is a darker diagonal, meaning that in all the communes, most of the people that spend their weekday noon in a given commune also live there.
The lightest element in the diagonal corresponds to Commune 1 (with 24\%, followed by Commune 2 with 43\%), because of the flow of people from the rest of the city that work there.

\subsection{Visualizing the City Pulse}
\label{sec:city_pulse}

\begin{figure}[th]
\vspace{-0.2cm}
\begin{minipage}{.44\textwidth}
  \centering
  \includegraphics[width=0.85\textwidth, clip=true, trim=58 40 10 20]{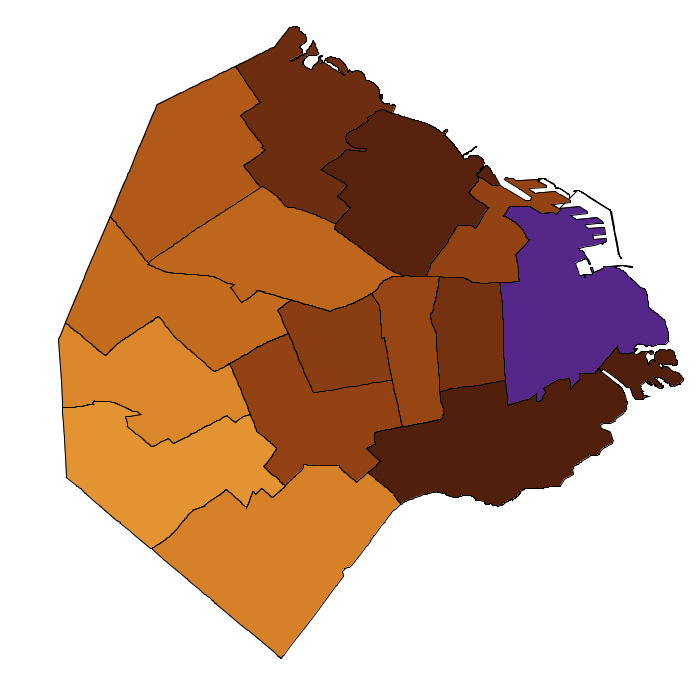}

  \small{(a)}
\end{minipage}
\begin{minipage}{.44\textwidth}
  \centering
  \includegraphics[width=0.85\textwidth, clip=true, trim=58 40 10 20]{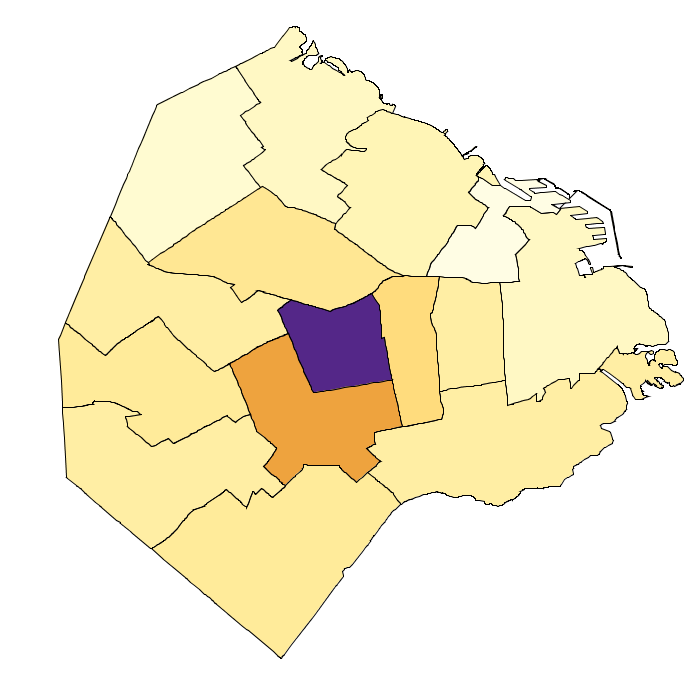}

  \small{(b)}
\end{minipage}
\begin{minipage}{.076\textwidth}
  \centering
  \includegraphics[width=0.90\textwidth]{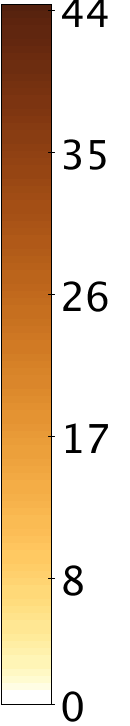}

  \phantom{\small{(c)}}
\end{minipage}
  \caption{\label{fig:visualization} Visualization of the number of people present on Monday to Thursday noon period in (a) Commune 1 and (b) Commune 6 (colored in violet) that live in each of the other communes. The scale shows the number of people (in thousands) each color represents.}
\label{fig:city_pulse_vis}
\end{figure}

Finally, Fig.~\ref{fig:city_pulse_vis} presents a visualization of the \emph{city pulse}. 
We plotted a map showing for Commune~1 and Commune~6 the number of people present there on a typical working day (Monday to Thursday) at noon, according to their home communes.
Commune 1 is the central business district so many people work there during the day, coming from very diverse locations. Commune 6, on the other hand, is one of the most populated and dense communes of the city (and represents its geographical center), but is mainly residential.
The difference in the number of people and variety of provenance between a central business district as Commune 1 and a more residential district as Commune 6 can be seen clearly in Fig.~\ref{fig:city_pulse_vis}.
We have also done a more complete analysis including other communes and day and hour groups (as shown in Table \ref{table:weekSlots}) achieving similar results.

\section{Conclusions and Future Work}

We presented a methodology to estimate the flow of people between different parts of the city
using mobile phone records.
According to our validation, the method is reliable, presenting
an average difference of 5\% with the origin-destination survey~\cite{enmodo2010}.

We extended the analysis to weekends using the proposed methodology, 
and found many interesting patterns which are coherent with our knowledge of the city. 
For instance, we showed how Commune 1, the central business district, yields during the weekends its role as major pole of attraction to Commune 14, which is a social and nightlife hub. 
We also presented a visualization where a business and a residential district can be clearly differentiated.
A more detailed analysis of this methodology was published in~\cite{anapolsky2014clatpu}.

We finally introduce ideas for future work:
(i) achieve a finer spatial granularity with a richer dataset;
(ii) consider the metropolitan region (suburbs) of the city in the analysis, as many people travel between the capital and its suburbs every day;
(iii) analyze the mobility of citizens during particular situations or events (for example, an evacuation or a holiday).

\bibliographystyle{unsrt}

\bibliography{mobility}

\end{document}